\documentclass[12pt]{article}
\usepackage{graphicx}
\usepackage{color}
\usepackage{amsmath,amssymb,amsfonts}
\usepackage{epsfig}
\usepackage{float}
\newtheorem{theorem}{Theorem}

\newtheorem{deff}{Definition}
\newcommand{\bq}{\begin{equation}}
\newcommand{\eq}{\end{equation}}
\newcommand{\bqa}{\begin{eqnarray}}
\newcommand{\eeqa}{\end{eqnarray}}
\begin{document}
\title{\vspace{-17mm}\rule{0mm}{10mm} \textbf{Barriers for the
reduction of transport due to the $E\times B$ drift in magnetized
plasmas}}
\author{Natalia Tronko $^*$
\and Michel Vittot $^*$
\and Cristel Chandre \thanks{Centre de Physique Th\'eorique (CPT) -
CNRS, Luminy - Case 907, 13288 Marseille cedex 09, France.\hspace{4cm}
Unit\'e Mixte de Recherche (UMR 6207) du CNRS, et des universit\'es
Aix-Marseille I, Aix-Marseille II et du Sud Toulon-Var. Laboratoire
affili\'e \`a la FRUMAM (FR 2291). Laboratoire de Recherche
Conventionn\'e du CEA (DSM-06-35).}
\and Philippe Ghendrih \thanks{Institut de Recherche sur la Fusion
Magn\'etique (IRFM/SCCP), CEA Cadarache, 13108 St-Paul-lez-Durance,
France.}
\and Guido Ciraolo \thanks{M2P2, UMR 6181, Technopole de Chateau
Gombert, 13451 Marseille Cedex 20, France.}
}
\date{2008 November 28}
\maketitle

\vspace{-10mm}

\begin{abstract}
We consider a $1\frac{1}{2}$ degrees of freedom Hamiltonian dynamical
system, which models the chaotic dynamics of charged test-particles in
a turbulent electric field, across the confining magnetic field in
controlled thermonuclear fusion devices. The external electric field
${\bf E} = -\,\nabla V$ is modeled by a phenomenological potential $V$
and the magnetic field ${\bf B}$ is considered uniform. It is shown
that, by introducing a small additive control term to the external
electric field, it is possible to create a transport barrier for this
dynamical system. The robustness of this control method is also
investigated. This theoretical study indicates that alternative
transport barriers can be triggered without requiring a control
action on the device scale as in present Internal Transport Barriers
(ITB).
\end{abstract}

\vspace{-7mm}

\tableofcontents

\vspace{7mm}
\section{Introduction}
It has long been recognised that the confinement properties of high
performance plasmas with magnetic confinement are governed by
electromagnetic turbulence that develops in microscales
\cite{Wagner1993}. In that framework various scenarios are explored to
lower the turbulent transport and therefore improve the overall
performance of a given device. The aim of such a research activity is
two-fold.

First, an improvement with respect to the basic turbulent scenario,
the so-called L-mode (L for low) allows one to reduce the reactor size
to achieve a given fusion power and to improve the economical
attractiveness of fusion energy production. This line of thought has
been privileged for ITER that considers the H-mode (H for high) to
achieve an energy amplification factor of $10$ in its reference
scenario \cite{ITER2007_chapter1}. The H-mode scenario is based on a
local reduction of the turbulent transport in a narrow regime in the
vicinity of the outermoster confinement surface \cite{Wagner1982}.

Second, in the so-called advanced tokamak scenarios, Internal
Transport Barriers are considered \cite{ITER2007_chapter1}. These
barriers are characterised by a local reduction of turbulent transport
with two important consequences, first an improvement of the core
fusion performance, second the generation of bootstrap current that
provides a means to generate the required plasma current in regime
with strong gradients \cite{ITER2007_chapter6}. The research on ITB
then appears to be important in the quest of steady state operation of
fusion reactors, an issue that also has important consequences for the
operation of fusion reactors.

While the H-mode appears as a spontaneous bifurcation of turbulent
transport properties in the edge plasma \cite{Wagner1982}, the ITB
scenarios are more difficult to generate in a controlled fashion
\cite{ITER2007_chapter2}. Indeed, they appear to be based on
macroscopic modifications of the confinement properties that are both
difficult to drive and difficult to control in order to optimise the
performance.

In this paper, we propose an alternative approach to transport
barriers based on a macroscopic control of the $E\times B$ turbulence.
Our theoretical study is based on a localized hamiltonian control
method that is well suited for $E\times B$ transport. In a previous
approach \cite{guido}, a more global scheme was proposed with a
reduction of turbulent transport at each point of the phase space. In
the present work, we derive an exact expression to govern a local
control at a chosen position in phase space. In principle, such an
approach allows one to generate the required transport barriers in the
regions of interest without enforcing large modification of the
confinement properties to achieve an ITB formation
\cite{ITER2007_chapter2}. Although the application of such a precise
control scheme remains to be assessed, our approach shows that local
control transport barriers can be generated without requiring
macroscopic changes of the plasma properties to trigger such barriers.
The scope of the present work is the theoretical demonstration of the
control scheme and consequently the possibility of generating
transport barriers based on more specific control schemes than
envisaged in present advanced scenarios.

In Section~\ref{sec:motivation}, we give the general description of
our model and the physical motivations for our investigation. In
Section~\ref{sec:control}, we explain the general method of {\it
localized control} for Hamiltonian systems and we estimate the size of
the control term. Section~\ref{sec:numerical} is devoted to the
numerical investigations of the control term, and we discuss its
robustness and its energy cost.
The last section~\ref{sec:conclusion} is devoted to conclusions and
discussion.
\section{\label{sec:motivation}Physical motivations and the $E\times
B$ model}
\subsection{Physical motivations}
Fusion plasma are sophisticated systems that combine the intrinsic
complexity of neutral fluid turbulence and the self-consistent
response of charged species, both electrons and ions, to magnetic
fields. Regarding magnetic confinement in a tokamak, a large external
magnetic field and a first order induced magnetic field are organised
to generate the so-called magnetic equilibrium of nested toroidal
magnetic surfaces \cite{Wesson2004}. On the latter, the plasma can be
sustained close to a local thermodynamical equilibrium. In order to
analyse turbulent transport we consider plasma perturbations of this
class of solutions with no evolution of the magnetic equilibrium, thus
excluding MHD instabilities. Such perturbations self-consistently
generate electromagnetic perturbations that feedback on the plasma
evolution. Following present experimental evidence, we shall assume
here that magnetic fluctuations have a negligible impact on turbulent
transport \cite{Fiksel1995}. We will thus concentrate on electrostatic
perturbations that correspond to the vanishing $\beta$ limit, where
$\beta=p/(B^2/2\mu_0)$ is the ratio of the plasma pressure $p$ to the
magnetic pressure. The appropriate framework for this turbulence is
the Vlasov equation in the gyrokinetic approximation associated to the
Maxwell-Gauss equation that relates the electric field to the charge
density. When considering the Ion Temperature Gradient instability
\cite{Garbet2007} that appears to dominate the ion heat transport, one
can further assume the electron response to be adiabatic so that the
plasma response is governed by the gyrokinetic Vlasov equation for the
ion species.

Let us now consider the linear response of such a distribution
function $\widehat{f}$, to a given electrostatic perturbation,
typically of the form $T_e\ \widehat{\phi}\ e^{-i\omega t + i
\vec{k}\vec{r}}$, (where $\widehat{f}$ and $\widehat{\phi}$ are
Fourier amplitudes of distribution function and electric potential).
To leading orders one then finds that the plasma response
exhibits a resonance:
\bq
\widehat{f} = \left(\frac{\omega + \omega^*}{\omega - k_{||} \ v_{||}}
-1 \right) \widehat{\phi} f_{eq}
\eq
Here $f_{eq}$ is the reference distribution function, locally
Maxwellian with respect to $v_{||}$ and $\omega^*$ is the diamagnetic
frequency that contains the density and temperature gradient that
drive the ITG instability \cite{Garbet2007}. $T_e$ is the electronic
temperature. This simplified plasma response to the electrostatic
perturbation allows one to illustrate the turbulent control that is
considered to trigger off transport barriers in present tokamak
experiments.

Let us examine the resonance $\omega -k_{||} \ v_{||}=0$ where $k_{||}=
(n-m/q)/R$ with $R$ being the major radius, $q$ the safety factor that
characterises the specific magnetic equilibrium and $m$ and $n$ the
wave numbers of the perturbation that yield the wave vectors of the
perturbation in the two periodic directions of the tokamak
equilibrium. When the turbulent frequency $\omega$ is small with
respect to $v_{th}/(qR)$, (where $v_{th}=\sqrt{k_B T/m}$ is the
thermal velocity), the resonance occurs for vanishing values of
$k_{||}$, and as a consequence at given radial location due to the
radial dependence of the safety factor. The resonant effect is
sketched on figure 1.
\begin{figure}[H]
\centering
\includegraphics[width=6cm]{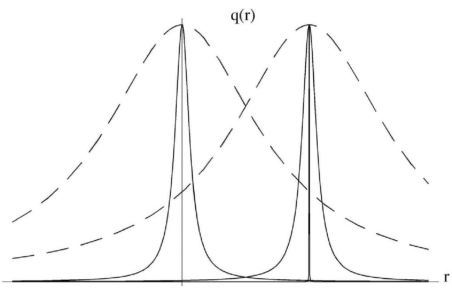}
\caption{Resonances for $q=\frac{m}{n}$ and $q=\frac{m+1}{n}$ for two
different widths, narrow resonances empedding large scale turbulent
transport and broad resonances favouring strong turbulent transport.}
\label{fig:resonance}
\end{figure}
In a quasilinear approach, the response to the perturbations will lead
to large scale turbulent transport when the width of the resonance
$\delta_m$ is comparable to the distance between the resonances
$\Delta_{m,m+1}$ leading to an overlap criterion that is comparable to
the well known Chirikov criterion for chaotic transport $\sigma_{m} =
(\delta_m + \delta_{m+1})/\Delta_{m,m+1}$ with $\sigma>1$ leading to
turbulent transport across the magnetic surfaces and $\sigma<1$
localising the turbulent transport to narrow radial regions in the
vicinity of the resonant magnetic surfaces.

The present control schemes are two-fold. First, one can consider a
large scale radial electric field that governs a Doppler shift of the
mode frequency $\omega$. As such the Doppler shift $\omega-\omega_E$
has no effect. However a shear of the Doppler frequency $\omega_E$,
$\omega_E=\bar{\omega}_E+\delta r \omega'_E$ will induce a shearing
effect of the turbulent eddies and thus control the radial extent of
the mode $\delta_m$, so that one can locally achieve $\sigma<1$ in
order to drive a transport barrier.

Second, one can modify the magnetic equilibrium so that the distance
between the resonant surfaces is strongly increased in particular in a
magnetic configuration with weak magnetic shear $(dq/dr\approx 0)$ so
that $\Delta_{m,m+1}$ is strongly increased, $\Delta_{m,m+1}\gg
\delta_m$, also leading to $\sigma<1$.

Both control schemes for the generation of ITBs can be interpreted
using the situation sketched on figure 1. The initial situation with
large scale radial transport across the magnetic surfaces (so called
L-mode) is indicated by the dashed lines and is governed by
significant overlap between the resonances. The ITB control scheme
aims at either reducing the width of the islands or increasing the
distance between the resonances yielding a situation sketeched by the
plain line in figure 1 where the overlap is too small and a region
with vanishing turbulent transport, the ITB, develops between the
resonances.

Experimental strategies in advanced scenarios comprising Internal
Transport Barriers are based on means to enforce these two control
schemes. In both cases they aim at modifying macroscopically the
discharge conditions to fulfill locally the $\sigma<1$ criterion. It
thus appears interesting to devise a control scheme based on a less
intrusive action that would allow one to modify the chaotic transport
locally by the choice of an appropriate electrostatic perturbation
hence leading to a local transport barrier.
\subsection{\label{subsec:EXB}The $E\times B$ model}
For fusion plasmas, the magnetic field $B$ is slowly variable with
respect to the inverse of the Larmor radius $\rho_L$ i.e: $\rho_L
|\nabla \ln B| \ll 1$. This fact allows the separation of the motion
of a charged test particle into a slow motion (parallel to the lines
of the magnetic field) and a fast motion (Larmor rotation). This fast
motion is named gyromotion, around some gyrocenter. In first
approximation the averaging of the gyromotion over the gyroangle gives
the approximate trajectory of the charged particle. This averaging is
the guiding-center approximation.

In this approximation, the equations of motion of a charged test
particle in the presence of a strong uniform magnetic field
$\mathbf{B}=B \hat{{\sf z}}$, (where $\hat{{\sf z}}$ is the unit
vector in the z direction) and of an external time-dependent electric
field $ \mathbf{E} = -\,\mathbf{\nabla} V_1 $ are:
\bqa
\frac{d}{dT}
\left(
\begin{array}{c}
X \\ \\ Y
\end{array}
\right)
& = &\frac{c \mathbf{E}\times \mathbf{B}}{B^2} =\frac{c}{B} \
\mathbf{E}(X,Y,T) \times \hat{{\sf z}} \nonumber \\
& = &
\frac{c}{B}
\left(
\begin{array}{c}
-\partial_{Y} V_1(X,Y,T)\\
\\
\partial_{X} V_1(X,Y,T)
\end{array}
\right)
\label{sys_0}
\eeqa
where $V_1$ is the electric potential. The spatial coordinates $X$ and
$Y$ play the role of canonically-conjugate variables and the electric
potential $V_1(X,Y,T)$ is the Hamiltonian for the problem. Now the
problem is placed into a parallelepipedic box with dimensions $L\times
\ell \times (2\pi/\omega)$, where $L$ and $\ell$ are some
characteristic lengths and $\omega$ is a characteristic frequency of
our problem, $X$ is locally a radial coordinate and $Y$ is a poloidal
coordinate. A phenomenological model \cite{Pettini} is chosen for the
potential:
\bq
V_1(X,Y,T)=\sum_{n,m=1}^N \frac{V_{0}\;\cos
\chi_{n,m}}{(n^2+m^2)^{3/2}}
\label{pot1}
\eq
where $V_0$ is some amplitude of the potential,
\[
\chi_{n,m} \equiv \frac{2\pi}{L} n X+ \frac{2\pi}{\ell} m
Y+\phi_{n,m}-\omega T
\]
$\omega$ is constant, for simplifying the numerical simulations and
$\phi_{n,m}$ are some random phases (uniformly distributed).

We introduce the dimensionless variables
\bq
(x, y, t) \equiv \left( 2\pi X/L,\; 2 \pi Y/\ell,\; \omega T \right)
\label{dimensionless_variables}
\eq
So the equations of motion (\ref{sys_0}) in these variables are:
\bq
\frac{d}{d{t}}
\left(
\begin{array}{c}
x\\ y
\end{array}
\right)
=
\left(
\begin{array}{c}
-\partial_{y} V(x,y,t)\\
\partial_{x} V(x,y,t)
\end{array}
\right)
\label{sys}
\eq
where $V= \varepsilon (V_1/V_0)$ is a dimensionless electric potential
given by
\bq
V(x,y,t)= \varepsilon\;\sum_{n,m=1}^N\frac{\cos\left(n x +m y
+\phi_{n,m}-t\right)}{(n^2+m^2)^{3/2}}
\label{pot}
\eq
Here
\bq
\varepsilon = 4\pi^2 (cV_0/B)/(L\ell\omega)
\label{epsilon}
\eq
is the small dimensionless parameter of our problem. We perturb the
model potential (\ref{pot}) in order to build a transport barrier. The
system modeled by Eqs.(\ref{sys}) is a $1 \frac{1}{2}$ degrees of
freedom system with a chaotic dynamics \cite{Pettini,guido}. The
poloidal section of our modeled tokamak is a Poincar\'e section for
this problem and the stroboscopic period will be chosen to be $2\pi$,
in term of the dimensionless variable $t$.

The particular choice (\ref{pot1}) or (\ref{pot}) is not crucial and
can be generalized. Generally, $\omega$ can be chosen depending on $n,
m$. This would make the numerical computations more involved. In the
following section, $V$ is chosen completely arbitrary.
\section{\label{sec:control}Localized control theory of hamiltonian
systems}
\subsection{\label{sec:control_term}The control term}
In this section we show how to construct a transport barrier for any
electric potential $V$. The electric potential $V(x,y,t)$ yields a
non-autonomous Hamiltonian. We expand the two-dimensional phase space
by including the canonically-conjugate variables ($E$,$\tau$),
\begin{equation}
H=H(E,x,y,\tau)=E+V(x,y,\tau)
\end{equation}
The Hamiltonian of our system thus becomes autonomous. Here $\tau$ is
a new variable whose dynamics is trivial:
$
\dot{\tau}=1 \ {\mathrm{i.e.}} \ \tau=\tau_0+t
$
and $E$ is the variable canonically conjugate to $\tau$. The Poisson
bracket in the expanded phase space for any $W=W(E,x,y,\tau)$ is given
by the expression:
\begin{equation}
\{W\} \equiv (\partial_x W)\partial_{y}-(\partial_{y}W)\partial_x +
(\partial_E W) \partial_{\tau}- (\partial_{\tau}W)\partial_{E}.
\label{bra}
\end{equation}
Hence $\{W\}$ is a linear (differential) operator acting on functions
of $(E,x,y,\tau)$. We call $H_0=E$ the unperturbed Hamiltonian and
$V(x,y,\tau)$ its perturbation. We now implement a perturbation theory
for $H_0$. The bracket (\ref{bra}) for the Hamiltonian $H$ is
\bq
\{H\}= (\partial_x V) \partial_{y} - (\partial_{y}V) \partial_x +
\partial_{\tau} - (\partial_{\tau} V) \partial_{E}
\eq
So the equations of motion in the expanded phase space are:
\bqa
\dot{y} & = & \{H\} y = \partial_{x}V(x,y,\tau) \\
\dot{x} & = & \{H\} x = -\,\partial_{y} V(x,y,\tau) \\
\dot{E} & = & \{H\}E = -\,\partial_{\tau} V(x,y,\tau)\\
\dot{\tau} & = & \{H\}\tau = 1
\eeqa
We want to construct a small modification $F$ of the potential $V$
such that
\bq
\widetilde{H} \equiv E+V(x,y,\tau)+F(x,y,\tau) \equiv
E+\widetilde{V}(x,y,\tau)
\eq
has a barrier at some chosen position $x=x_0$.
So the control term
\bq
F=\widetilde{V}(x,y,\tau)-V(x,y,\tau)
\label{contr}
\eq
must be much smaller than the perturbation (e.g., quadratic in $V$).
One of the possibilities is:
\bq
\widetilde{V}\equiv V(x+\partial_{y}f(y,\tau),y,\tau)
\label{potc}
\eq
where
\[
f(y,\tau)\equiv\int_0^{\tau} V(x_0,y,t) dt
\]
Indeed we have the following theorem:
\begin{theorem}
The Hamiltonian $\widetilde{H}$ has a trajectory $x=x_0+\partial_y
f(y,\tau)$ acting as a barrier in phase space.
\end{theorem}
{\textbf{Proof}} \\
Let the Hamiltonian $\widehat{H} \equiv \exp(\{f\})\tilde{H}$ be
canonically related to $\widetilde{H}$. (Indeed the exponential of any
Poisson bracket is a canonical transformation.) We show that
$\widehat{H}$ has a simple barrier at $x=x_0$. We start with the
computation of the bracket (\ref{bra}) for the function $f$. Since
$f=f(y,\tau)$, the expression for this bracket contains only two
terms,
\bq
\{f\}\ \equiv\ -f'\partial_x \ -\ \dot{f} \partial_E
\eq
where
\bq
f'\equiv \partial_{y}f \ {\mathrm{and}} \ \dot{f}
\equiv\partial_{\tau}f
\eq
which commute:
\bq
[f'\partial_x ,\dot{f} \partial_E]=0
\label{commut}
\eq
Now let us compute the coordinate transformation generated by
$\exp(\{f\})$:
\bq
\exp(\{f\}) \;\equiv\; \exp(-f'\partial_x)\;\exp(- \dot{f}
\partial_E),
\eq
where we used (\ref{commut}) to separate the two exponentials.

Using the fact that $\exp(b\partial_x)$ is the translation operator of
the variable $x$ by the quantity $b$: $[\exp(b \partial_x)W](x) =
W(x+b)$, we obtain
\bqa
\widehat{H} & = & e^{\{f\}}{\tilde H} \;\equiv\; e^{\{f\}}E \;+\;
e^{\{f\}}\tilde{V}(x,y,\tau) \nonumber \\
& = & \left( E \;-\; \dot{f} \right) \;+\;
\widetilde{V}\left(x-f^{\prime}, y, \tau\right) \nonumber \\
& = & E \;-\; V(x_0,y,\tau) \;+\; V(x+f'-f',y,\tau) \nonumber \\
& = & E-V(x_0,y,\tau)+ V(x,y,\tau)
\eeqa
This Hamiltonian has a simple trajectory $x=x_0, E=E_0$, i.e. any
initial data $x=x_0, \ y=y_0,E=E_0,\tau=\tau_0$ evolves under the flow
of $\widehat{H}$ into $x=x_0,y=y_t,E=E_0,\tau=\tau_0+t$ for some
evolution $y_t$ that may be complicated, but not useful for our
problem. Hamilton's equations for $x$ and $E$ are now
\bqa
\dot{x}&=&\{\widehat{H}\}x=\partial_{y}\left[V(x_0,y,\tau)-V(x,
y,\tau)\right] \\
\dot{E}&=&\{\widehat{H}\}E = \partial_{\tau} \left[V(x_0,y,\tau) -
V(x,y,\tau) \right]
\eeqa
so that for $x = x_{0}$, we find $\dot{x} = 0 = \dot{E}$. Then the
union of all points $(x, y, E, \tau)$ at $x = x_{0}\ \ E = E_{0}$:
\bq
\mathfrak{B}_0=\bigcup_
{y,\tau,E_0}
\left(
\begin{array}{c}
x_0 \\ y \\ E_0 \\ \tau
\end{array}
\right)
\eq
is a $3$-dimensional surface ${\mathbb{T}}^2\times{\mathbb{R}}$,
(${\mathbb{T}}\equiv{\mathbb{R}}/2\pi{\mathbb{Z}}$) preserved by the
flow of $\widehat{H}$ in the $4$-dimensional phase space. If an
initial condition starts on ${\mathfrak{B}}_0$, its evolution under
the flow $\exp(t \{\hat{H}\})$ will remain on ${\mathfrak{B}}_0$.

So we can say that $\mathfrak{B}_0$ act as a barrier for the
Hamiltonian $\widehat{H}$: the initial conditions starting inside
$\mathfrak{B}_0$ can't evolve outside $\mathfrak{B}_0$ and vice-versa.

To obtain the expression for a barrier $\mathfrak{B}$ for
$\widetilde{H}$ we deform the barrier for $\widehat{H}$ via the
transformation $\exp(\{f\})$. As
\bq
\widetilde{H}=e^{-\{f\}}\widehat{H}
\eq
and $\exp(\{f\})$ is a canonical transformation, we have
\bq
\{\widetilde{H}\}= \{e^{-\{f\}} \widehat{H}\} = e^{-\{f\}}
\{\widehat{H}\} e^{\{f\}}
\eq
Now let us calculate the flow of $\widetilde{H}$:
\bq
e^{t\{\widetilde{H}\}} = e^{t\left(e^{-\{f\}} \{\widehat{H}\}
{e^{\{f\}}}\right)} = e^{-\{f\}} e^{t\{\widehat{H}\}} {e^{\{f\}}}
\label{flot}
\eq
Indeed:
\bq
e^{t\left(e^{-\{f\}} \{\widehat{H}\} {e^{\{f\}}}\right)} =
\sum_{n=0}^{\infty} \frac{t^n(e^{-\{f\}} \{\widehat{H}\}
e^{\{f\}})^n}{n!}
\eq
For instance when $n=2$:
\bqa
t^2(e^{-\{f\}}\{\widehat{H}\}e^{\{f\}})^2 & = & t^2
e^{-\{f\}}\{\widehat{H}\}e^{\{f\}}e^{-\{f\}}\{\widehat{H}\}e^{\{f\}}
\nonumber \\
& = & t^2 e^{-\{f\}}{\{\widehat{H}\}}^2 e^{\{f\}}
\eeqa
and so
\bq
e^{t\{\widetilde{H}\}}=\sum_{n=0}^{\infty}\frac{t^n
e^{-\{f\}}\{\widehat{H}\}^n
e^{\{f\}}}{n!}=e^{-\{f\}}e^{t\{\widehat{H}\}}{e^{\{f\}}}
\eq
As we have seen before:
\[ e^{\{f\}} \left(
\begin{array}{c}
x \\ y \\ E \\ \tau
\end{array}
\right)=
\left(
\begin{array}{c}
x-f' \\ y \\ E -\dot{f} \\ \tau
\end{array}
\right) \]
and
\bq
e^{t\{\widehat{H}\}}
\left(
\begin{array}{c}
x_0\\ y \\ E_0 \\ \tau
\end{array}
\right)=
\left(
\begin{array}{c}
x_0 \\ y_t \\ E_0 \\ \tau+t
\end{array}
\right)
\eq
Multiplying (\ref{flot}) on the right by $e^{-\{f\}}$ we obtain:
\[
e^{t\{\widetilde{H}\}}e^{-\{f\}} \;=\; e^{-\{f\}}e^{t\{\widehat{H}\}}
\]
\bq
e^{t\{\widetilde{H}\}}e^{-\{f\}}
\left(
\begin{array}{c}
x_0 \\ y \\ E_0 \\ \tau
\end{array}
\right)=
e^{t\{\widetilde{H}\}}
\left(
\begin{array}{c}
x_0+f'(y,\tau) \\ y \\ E_0+\dot{f}(y,\tau) \\ \tau
\end{array}
\right)
\eq
and
\bqa
e^{-\{f\}}e^{t\{\widehat{H}\}}
\left(
\begin{array}{c}
x_0 \\ y \\ E_0 \\ \tau
\end{array}
\right) & = &
e^{-\{f\}}
\left(
\begin{array}{c}
x_0 \\ y_t \\ E_0 \\ \tau +t
\end{array}
\right) \nonumber \\
& = & \left(
\begin{array}{c}
x_0+f'(y_t,\tau+t) \\ y_t \\ E_0+\dot{f}(y_t,\tau+t) \\ \tau+t
\end{array}
\right)
\eeqa
So the flow $\exp(t\{\widetilde{H}\})$ preserves the set
\bq
\mathfrak{B}=
\bigcup_{y,\tau,E_0}
\left(
\begin{array}{c}
x_0+f'(y,\tau) \\ y \\ E_0+\dot{f}(y,\tau) \\ \tau
\end{array}
\right)
\eq
$\mathfrak{B}$ is a $3$ dimensional invariant surface, topologically
equivalent to $\mathbb{T}^2\times \mathbb{R}$ into the $4$ dimensional
phase space. $\mathfrak{B}$ separates the phase space into 2 parts,
and is a barrier between its interior and its exterior. $\mathfrak{B}$
is given by the deformation $\exp(\{f\})$ of the simple barrier
$\mathfrak{B}_0$.

The section of this barrier on the sub space $(x,y,t)$ is
topologically equivalent to a torus $\mathbb{T}^2$.

\vspace{10mm}
This method of control has been successfully applied to a real
machine: a traveling wave tube to reduce its chaos \cite{Doveil}.
\subsection{Properties of the control term}
In this Section, we estimate the size and the regularity of the
control term (\ref{contr}).
\begin{theorem}
For the phenomenological potential (\ref{pot}) the control term
(\ref{contr}) verifies:
\bq
\|F\|_{\frac{1}{N},\frac{1}{N}}\leq \varepsilon^2 N^2 \frac{e^3}{4
\pi}
\label{est1}
\eq
if $\varepsilon$ is small enough, i.e. if
$
|\varepsilon|\leq\frac{\sqrt{\pi}}{2 N e^{3/2}}
$
where $N$ is the number of modes in the sum (\ref{pot}).
\end{theorem}

\textbf{Proof}
The proof of this estimation is given in \cite{math} and is based on
rewriting
\bqa
F=V(x+f')-V(x) & = & \int_0^1 ds \ \partial_x V(x+s f', y ,\tau)
f'(y,\tau) \nonumber \\
& = & {\mathcal{O}}(V^2)
\eeqa
and then use Cauchy's Theorem.
\section{\label{sec:numerical}Numerical investigations for the control
term}
\begin{figure}[H]
\centering
\includegraphics[width=\textwidth]{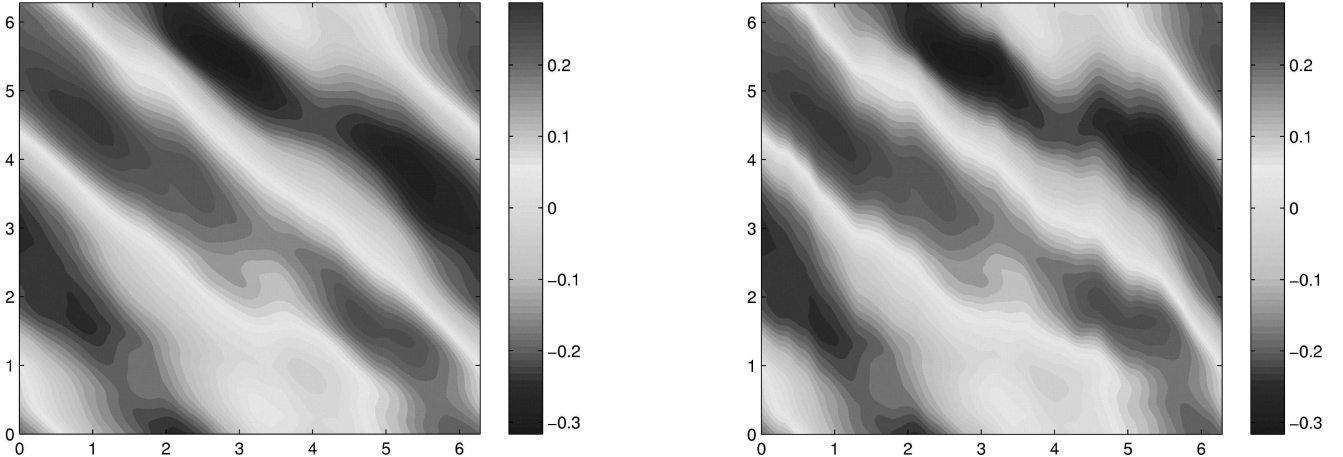}
\caption{Uncontrolled and controlled potential for $\varepsilon=0.6$,
$t=\frac{\pi}{4}$, $x_0=2$.}
\label{fig:fig_1}
\end{figure}

In this Section, we present the results of our numerical
investigations for the control term $F$. The theoretical estimate
presented in the previous section shows that its size is quadratic in
the perturbation. Figure \ref{fig:fig_1} shows the contour plot of
$V(x,y,t)$ and $\widetilde{V}(x,y,t)$ ($\widetilde{V}=V+F$) at some
fixed time $t$, for example $t=\frac{\pi}{4}$. One can see that the
contours of both potentials are very similar. But the dynamics of the
systems with $V$ and $\tilde{V}$ are very different.

For all numerical simulations we choose the number of modes $N=25$ in
(\ref{pot}). In all plots the abscissa is $x$ and the ordinate is $y$.
\subsection{Phase portrait for the exact control term}
\begin{figure}[H]
\centering
\includegraphics[width=\textwidth]{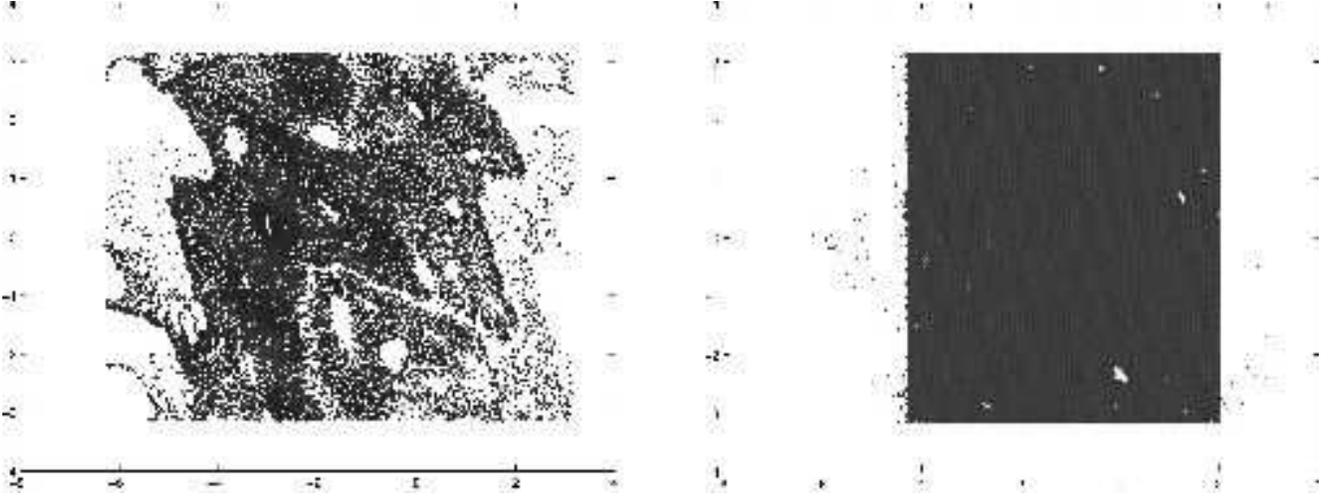}
\caption{Phase portraits without control term and with the exact
control term, for $\varepsilon=0.9$, $x_0=2$, $N_{traj}=200$.}
\label{fig:fig_2}
\end{figure}

To explore the effectiveness of the barrier, we plot (in
Fig.~\ref{fig:fig_2}) the phase portraits for the original system
(without control term) and for the system with the exact control term
$F$. We choose the same initial conditions. The time of integration is
$T=2000$, the number of trajectories: $N_{traj}=200$ (number of
initial conditions, all taken in the strip $-1-\pi\leq x \leq -\pi$;
$0\leq y \leq2\pi$) and the parameter $\varepsilon=0.9$. We choose the
barrier at position $x_0=2$. And to get a Poincar\'e section, we plot
the poloidal section when $t \in 2 \pi{\mathbb{Z}}$. Then we compare
the number of trajectories passing through the barrier during this
time of integration for each system. We eliminate the points after the
crossing. For the uncontrolled system $68\%$ of the initial conditions
cross the barrier at $x_0=2$ and for the controlled system only $1\%$
of the trajectories escape from the zone of confinement. The theory
announces the existence of an exact barrier for the controlled system:
these escaped trajectories ($1\%$) are due to numerical errors in the
integration.
One can observe that the barrier for the controlled system is a
straight line. In fact this barrier moves, its expression depends on
time:
\bq
x=x_0+f'(y,t)
\eq
But when $t \in 2 \pi{\mathbb{Z}}$ its oscillation around $x=x_0$
vanishes: $f'(y,2k\pi)=\int_0^{2 k \pi} \partial_y V(x_0,y,t)dt=0$.
This is what we see on this phase portrait. In fact we create $2$
barriers at position $x=x_0$, and $x=x_0-2 \pi$ (and also at $x_0+2
n\pi$) because of the periodicity of the problem. We note that the
mixing increases inside the two barriers. The same phenomenon was also
observed in the control of fluids \cite{fluid}, where the same method
was applied.
\subsection{Robustness of the barrier}
In a real Tokamak, it is impossible to know an analytical expression
for electric potential $V$. So we can't implement the exact expression
for $F$. Hence we need to test the robustness of the barrier by
truncating the Fourier decomposition (for instance in time) of the
controlled potential.
\subsection*{Fourier decomposition}
\begin{theorem}
The potential (\ref{potc}) can be decomposed as $\widetilde{V} =
\sum_{k\in{\mathbb{Z}}} \widetilde{V}_k$, where
\bq
\widetilde{V}_k = \varepsilon \sum_{n,m=1}^N
\frac{{\mathcal{J}}_k(n\rho)}{(n^2+m^2)^{3/2}}
\cos\left(\eta+k\Theta+(k-1)t\right) \label{vk}\\
\eq
with
\bqa
\eta_{n,m}(y)&=&n x +m y+\phi_{n,m} + n \varepsilon F_c \\
F_c(y) &=& \sum_{n,m=1}^N \frac{m\;\cos(K_{n,m,y})} {(n^2+m^2)^{3/2}}
\label{Fc}\\
F_s(y) &=& \sum_{n,m=1}^N \frac{m\; \sin(K_{n,m,y})} {(n^2+m^2)^{3/2}}
\label{Fs}\\
K_{m,n,y}&=&nx_0+my+\phi_{n,m}
\eeqa
and ${\mathcal{J}}_k$ is the Bessel's function
\bq
{\mathcal{J}}_k(n\rho)=\frac{1}{\pi} \int_0^{\pi}
\cos\left(ku-n\rho\sin u\right) du
\eq
\end{theorem}

\textbf{Proof}
We rewrite explicitly the expression (\ref{potc}) for our
phenomenological controlled potential $\widetilde{V}(x,y,t)$:
\bq
\widetilde{V}(x,y,t) = \!\varepsilon \!\!\!\sum_{n,m=1}^N
\!\!\frac{\cos\Bigl(n (x + f'(y,t)) + m y + \phi_{n,m} - t \Bigr)}
{(n^2+m^2)^{3/2}}
\eq
with
\bq
f'(y,t) = \varepsilon \sum_{n,m=1}^N\frac{m\ \Bigl(\cos K_{n,m,y} -
\cos(K_{n,m,y}-t)\Bigr)}{(n^2+m^2)^{3/2}}
\eq
With the definition (\ref{Fc}) and (\ref{Fs}) we have:
\bq
f'(y,t) = \varepsilon(F_c(y)\ (1 - \cos t) - F_s(y)\ \sin t)
\eq
Let us introduce
\bq
\rho =\varepsilon (F_c^2 + F_s^2)^{1/2}
\eq
and $\Theta$ by
\bq
\rho\sin\Theta \equiv -\varepsilon F_c(y) \hspace{15mm}
\rho\cos\Theta \equiv -\varepsilon F_s(y)
\eq
so that
\bq
\widetilde{V} = \varepsilon \sum_{n,m=1}^N
\frac{\cos\left(\eta-t +
n\rho\sin(\Theta+t) \right)}{(n^2+m^2)^{3/2}}
\eq
Using Bessel's functions properties \cite{Abramovich}
\bq
\cos(\rho\sin\Theta) = \sum_{k\in{\mathbb{Z}}}{\mathcal{J}}_k(\rho)
\cos k\Theta
\eq
\bq
\sin(\rho\sin\Theta) = \sum_{k\in{\mathbb{Z}}}{\mathcal{J}}_k(\rho)
\sin k\Theta
\eq
we get
\bq
\cos\left(\eta- t + n\rho\sin(\Theta+t) \right) =
\sum_{k\in{\mathbb{Z}}} {\mathcal{J}_k(n\rho) \cos\left(\xi\right)}
\eq
where $\xi=\eta+k\Theta+(k-1)t$,
and we finally obtain (\ref{vk}). The theorem is proved. $\blacksquare$

During numerical simulations we truncate the controlled potential by
keeping only its first $3$ temporal Fourier's harmonics:
\bq
\widetilde{V}_{tr}\!\! =\! \varepsilon \!\!\!\! \sum_{n,m=1}^N \!\!\!
\frac{A_0 + A_1 \cos t + B_1 \sin t + A_2\cos 2t + B_2\sin 2t}
{(n^2+m^2)^{3/2}}
\label{potr}
\eq
\bqa
&&A_0={\mathcal{J}}_{0}(n\rho)\cos(\eta+\Theta) \nonumber\\
&&A_1={\mathcal{J}}_{0}(n\rho)\cos\eta+{\mathcal{J}}_{2}(n\rho)
\cos(\eta+2\Theta)\nonumber\\
&&B_1={\mathcal{J}}_{0}(n\rho)\sin\eta-{\mathcal{J}}_{2}(n\rho)
\sin(\eta+2\Theta)\nonumber
\\
&&A_2={\mathcal{J}}_{3}(n\rho)\cos(\eta+3\Theta)-
{\mathcal{J}}_{1}(n\rho)\cos(\eta-\Theta)\nonumber\\
&&B_2=-{\mathcal{J}}_{3}(n\rho)\sin(\eta+3\Theta)-
{\mathcal{J}}_{1}(n\rho)\sin(\eta-\Theta)\nonumber
\eeqa
\begin{figure}[H]
\centering
\includegraphics[width=\textwidth]{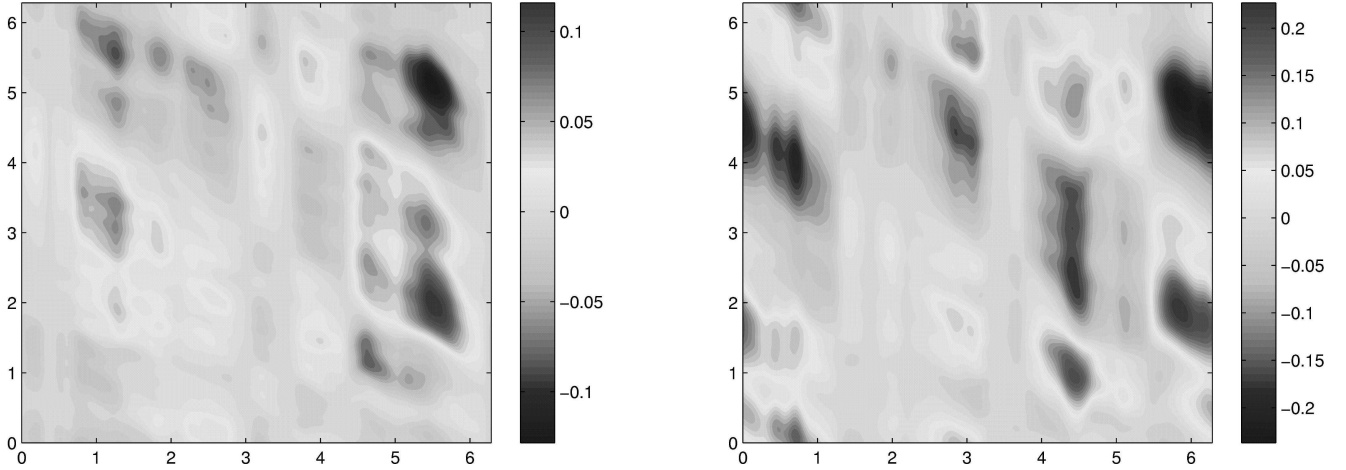}
\caption{Exact Control Term and Truncated Control Term with
$\varepsilon=0.6, t=\frac{\pi}{4} $.}
\label{fig:4}
\end{figure}
Figure \ref{fig:4} compares the two contour plots for the exact
control term and the truncated control term (\ref{potr}).
Figure\ref{fig:5} compares the two phase portraits for the system
without control term and for the system with the above truncated
control term (\ref{potr}). The computation of $\widetilde{V}_{tr}$ on
some grid has been performed in {\sf Matlab} and the numerical
integration of the trajectories was done in {\sf C}.
\begin{figure}[H]
\centering
\includegraphics[width=\textwidth]{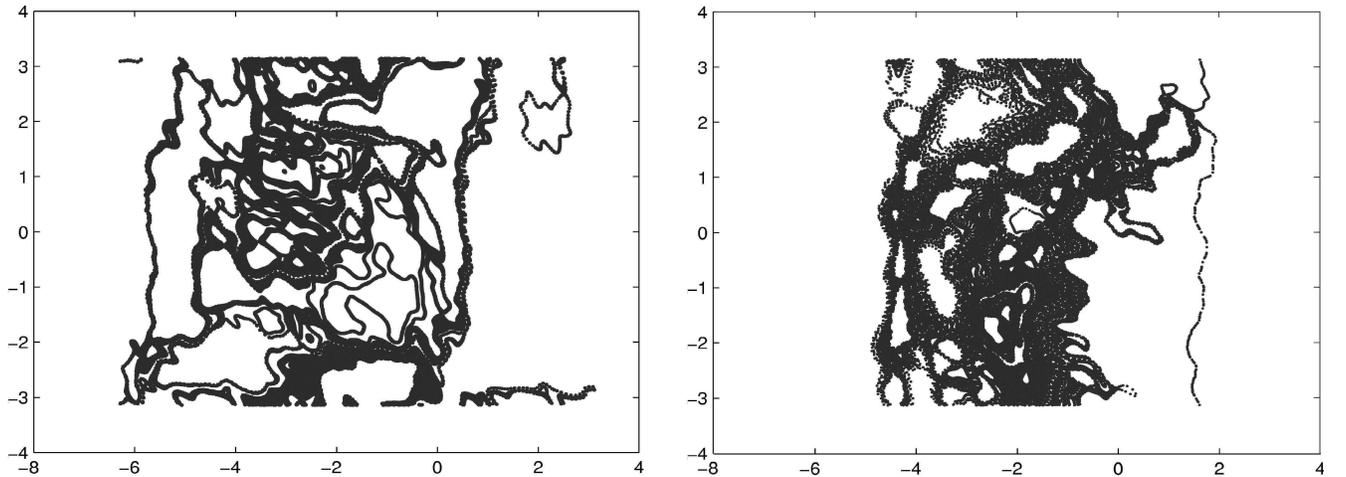}
\caption{$\varepsilon=0.3$, $T=2000$, $N_{traj}=50$.}
\label{fig:5}
\end{figure}
One can see a barrier for the system with the truncated control term.
As for the system with the exact control term we create two barriers
at positions $x=x_0$ and $x=x_0-2 \pi$ and the phenomenon of
increasing the mixing inside the barriers persist.
\subsection{Energetical cost}
As we have seen before, the introduction of the control term into the
system can reduce and even stop the diffusion of the particles through
the barrier. Now we estimate the energy cost of the control term $F$
and the truncated control term $F_{tr}\equiv \widetilde{V}_{tr}-V$.
\begin{deff}
The average of any function $W=W(x,y,\tau)$ is defined by the formula:
\bq
<|W|>=\int_0^{2 \pi}\!\!dx\!\!\int_0^{2 \pi}\!\! dy\!\!\int_0^{2
\pi}\!\! dt \ |W(x,y,t)|
\eq
\end{deff}
Now we calculate the ratio between the absolute value of the truncated
control (electric potential) or the exact control and the uncontrolled
electric potential:
\[
\zeta_{ex} = <|F|^2>/<|V|^2>
\]
and
\[
\zeta_{tr}= <|F_{tr}|^2>/<|V|^2>
\]
We also compute the ratio between the energy of the control electric
field and the energy of the uncontrolled system in their exact and
truncated version
\[
\eta_{ex}= <|\nabla{F}|^2>/<|\nabla{V}|^2>
\]
and
\[
\eta_{tr}= <|\nabla{F}_{tr}|^2>/<|\nabla{V}|^2>
\]
for different values of $\varepsilon$. Results are shown in Table
\ref{tab:Table_I}.

\begin{table}
\caption{\label{tab:Table_I}Squared ratios of the amplitudes of the
control term and the uncontrolled electric potential
$\zeta_{ex},\zeta_{tr}$; ratios of electric energy of the control term
and the uncontrolled electic potential $\eta_{ex},\eta_{tr}$; for the
system with exact and truncated control term.}
\begin{tabular}{ccccc}
$\varepsilon$ & $\zeta_{ex}$ & $\zeta_{tr}$ & $\eta_{ex}$ &
$\eta_{tr}$\\ \hline
$0.3$ &$0.1105$ &$0.1193$ &$0.6297$ &$0.1431$ \\
$0.4$ &$0.1466$ &$0.1583$ &$0.7145$ &$0.2393$ \\
$0.5$ &$0.1822$ &$0.1967$ &$0.8161$ &$0.3550$ \\
$0.6$ &$0.2345$ &$0.2137$ &$0.9336$ &$0.4883$ \\
$0.7$ &$0.2518$ &$0.2716$ &$1.0657$ &$0.6375$ \\
$0.8$ &$0.2858$ &$0.3038$ &$1.2119$ &$0.8014$ \\
$0.9$ &$0.3191$ &$0.3439$ &$1.3722$ &$0.9796$ \\
$1.5$ &$0.5052$ &$0.5427$ &$2.6247$ &$2.3037$ \\ \hline
\end{tabular}
\end{table}

One can see that the truncated control term needs a smaller energy
than the exact control term. In Table \ref{tab:Table_II}, we present
the number of particles passing through the barrier in function of
$\varepsilon$, after the same integration time.

\begin{table}
\caption{\label{tab:Table_II}Number of escaping particles without
control term ${\mathcal{N}}_{without}$, and for the system with the
exact control term ${\mathcal{N}}_{exact}$ and the truncated control
term ${\mathcal{N}}_{tr}$.}
\begin{tabular}{cccc}
$\varepsilon$ & ${\mathcal{N}}_{without}$ & ${\mathcal{N}}_{exact}$ &
${\mathcal{N}}_{tr}$ \\ \hline
$0.4$ & $22\%$ &$0\%$ &$6 \%$ \\
$0.5$ & $26\%$ &$0\%$ &$18\%$ \\
$0.9$ & $68\%$ &$1\%$ &$44\%$ \\
$1.5$ & $72\%$ &$1\%$ &$54\%$ \\ \hline
\end{tabular}
\end{table}

Let $\Delta{\mathcal{N}}= {\mathcal{N}}_{without}-{\mathcal{N}}_{tr}$
be the difference between the number of particles passing through the
barrier for the system without control and with the truncated control
and $\Delta\eta=\eta_{ex}-\eta_{tr}$ the difference between the
relative electric energy for the system with the exact control term
and the system with the truncated control term. In Table
\ref{tab:Table_III} we present $\Delta{\mathcal N} $ and $\Delta \eta$
for differents values of $\varepsilon$.

\begin{table}
\caption{\label{tab:Table_III}Difference $\Delta \mathcal{N}$ of the
number of particles passing trough the barrier and difference of
relative electric energy
$\Delta \eta$ for the controlled and uncontrolled system.}
\begin{tabular}{ccc}
$\varepsilon$ & $\Delta{\mathcal{N}}$ & $\Delta \eta $ \\ \hline
$0.3$ & $8 \%$ & $0.49$ \\
$0.4$ & $16\%$ & $0.47$ \\
$0.5$ & $8 \%$ & $0.46$ \\
$0.9$ & $24\%$ & $0.39$ \\
$1.5$ & $18\%$ & $0.32$ \\ \hline
\end{tabular}
\end{table}

For $\varepsilon$ below $0.2$ the non controlled system is rather
regular, there is no particles stream through the barrier, so we have
no need to introduce the control electric field. For $\varepsilon$
between $0.3$ and $0.9$ the truncated control field is quite
efficient, it allows to drop the chaotic transport through the barrier
by a factor $8 \% $ to $24 \% $ with respect to the uncontrolled
system and it requires less energy than the exact control field. For
$\varepsilon$ greater than $1$ the truncated control field is less
efficient than the exact one, because the dynamics of the system is
very chaotic. For example when $\varepsilon=1.5$, there are $72\% $ of
the particles crossing the barrier for the uncontrolled system and
$54\% $ for the system with the truncated control field. At the same
time the energetical cost of the truncated control field is above
$70\% $ of the exact one, which allows to stop the transport through
the barrier. So for $\varepsilon\geq 1$ we need to use the exact
control field rather than the truncated one.
\section{\label{sec:conclusion}Discussion and Conclusion}
In this article, we studied a possible improvement of the confinement
properties of a magnetized fusion plasma. A transport barrier
conception method is proposed as an alternative to presently achieved
barriers such as the H-mode and the ITB scenarios.
One can remark, that our method differs from an ITB construction.
Indeed, in order to build-up a transport barrier, we do not require a
hard modification of the system, such as a change in the q-profile.
Rather, we propose a weak change of the system properties that allow a
barrier to develop. However, our control scheme requires some
knowledge and information relative to the turbulence at work, these
having weak or no impact on the ITB scenarios.
\subsection{Main results}
First of all we have proved that the local control theory gives the
possibility to construct a transport barrier at any chosen position
$x=x_0$ for any electric potential $V(x,y,t)$. Indeed, the proof given
in section~\ref{sec:control} does not depend on the model for the
electric potential $V$. In Subsection~\ref{sec:control_term}, we give
a rigorous estimate for the norm of the control term $F$, for some
phenomenological model of the electric potential. The introduction of
the exact control term into the system inhibits the particle transport
through the barrier for any $\varepsilon$ while the implementation of
a truncated control term reduces the particle transport significantly
for $\varepsilon \in (0.3,1.0)$.
\subsection{Discussion, open questions}
\subsubsection{Comparison with the global control method}
Let us now compare our approach with the global control method
\cite{guido} which aims at globally reducing the transport in every
point of the phase space. Our approach aims at implementing a
transport barrier. However, one also observes a global modification of
the dynamics since the mixing properties seem to increase away from
the barriers.

Furthermore, in many cases, only the first few terms of the expansion
of the global control term \cite{guido} can be computed explicitly.
Here we have an explicit exact expression for the local control term.
\subsubsection{Effectiveness and properties of the control procedure}
In subsection \ref{subsec:EXB}, we have introduced the dimensionless
variables (\ref{dimensionless_variables}) and defined a dimensionless
control parameter $\varepsilon\equiv 4\pi^2 (cV_0/B)/(L\ell\omega)$.
In the simplifying case where $l= L = 2\pi/k$ is the characteristic
length of our problem, we have $\varepsilon=c k^2 V_0/(\omega B)$. Let
us consider a symmetric vortex, hence with characteristic scale $1/k$.
Let us now consider the motion of a particle governed by such a
vortex. The order of magnitude of the drift velocity is therefore $v_E
= k c V_0/B$ and the associated characteristic time $\tau_{ETT}$,
$\tau_{ETT}\equiv 1/ (k v_E)$, is the eddy turn over time. Let
$\omega$ be the characteristic evolution frquency of the turbulent
eddies, here of the electric field, then the Kubo number $K$ is
$K=1/\omega \tau_{ETT}$. This parameter is the dimensionless control
parameter of this class of problems, and we remark that in our case $K
=\varepsilon$. It is also important to remark that the parameter $K$
also characterises the diffusion properties of our system. Indeed, let
$\delta$ be a step size of our particle in a random walk process and
let $\tau$ be the associated characteristic time, the diffusion
coefficient is then $D=\delta^2/\tau$. Since one can relate the
characteristic step and time by the velocity, $\delta=v_E \tau$, on
also finds:
\bq
D=\frac{(v_E \tau)^2}{\tau}=\frac{k^2 c^2 V_0^2}{ B^2}
\tau=\frac{1}{k^2 \tau_{ETT}^2}\tau=\frac{K^2}{k^2}\omega^2\tau
\eq
We also introduce the reference diffusion coefficient
$\bar{D}=k^{-2}\omega$, so that:
\bq
D / \bar{D} \equiv K^2 \omega \tau
\label{diffusion}
\eq
They are two asymptotic regimes for our system. The first one, is the
regime of weak turbulence, characterised by $\omega \tau_{ETT} \gg 1$
and therefore $K \ll 1$. In this regime, the electric potential
evolution is fast, the particle trajectories only follow the eddy
geometry on distances much smaller than the eddy size. The steps
$\delta$ are small and the characteristic time $\tau$ of the random
walk such that $\omega \tau \approx 1$. The particle diffusion
(\ref{diffusion}) is then such that:
\bq
D /\bar{D}\approx K^2~~~~for~~~~~~\omega \tau_{ETT} \gg 1
\eq
The second asymptotic regime is the regime of strong turbulence, with
$\omega \tau_{ETT} \ll 1$ and $K \gg 1$. Particles then explore the
eddies before decorrelation and the characteristic time of the random
step is typically $\tau \approx \tau_{ETT}$ and:
\bq
D /\bar{D}\approx K~~~~for~~~~~~\omega \tau_{ETT} \ll 1
\eq
The first regime corresponds to the weak turbulence limit with weak
Kubo number and particle diffusion and the second to strong turbulence
and large Kubo number and particle diffusion. The control method
developed in this article does not depend on $K\equiv\varepsilon$.
There is always a possibility to construct an exact transport barrier.
However for the numerical simulations, we have remarked, that for
small $\varepsilon$ one can observe a stable barrier without escaping
particles, and for $\varepsilon$ close or more than $1$ there is some
leaking of particles across the barrier. The barrier is more difficult
to enforce. Also when considering the truncated control term, one
finds that the control term is ineffective in the strong turbulence
limit.

Let us now consider the implementation of our method to turbulent
plasmas where the turbulent electric field is consistent with the
particle transport. The theoretical proof of an hamiltonian control
concept is developped provided the system properties at work are
completely known. For example the analytic expression for the electric
potential. This is impossible in a real system, since the measurements
take place on a finite spatio-temporal grid. This has motivated our
investigation of the truncated control term by reducing the actually
used information on the system. As pointed out previously, one finds
that this approach is ineffective for strong turbulence. Another issue
is the evolution of the turbulent electric field following the
appearance of a transport barrier. This issue would deserve a specific
analysis and very likely updating the control term on a trasnport
characteristic time scale. An alternative to such a process would be to
use a retroactive Hamiltonian approach (a classical field theory)
\cite{Birula} and to develop the control theory in that framework.
\vspace{7mm}

{\large\bf{Acknowledgements}}

We acknowledge very useful and encouraging discussions with A.
Brizard, M. Vlad and M. Pettini. This work supported by the European
Communities under the contract of Association between EURATOM and CEA
was carried out within the framework of the European Fusion
Development Agreement. The views and opinions expressed herein do not
necessarily reflect those of the European Commission.

\end{document}